\documentclass[epj,nopacs]{svjour}
\usepackage{amsfonts}
\usepackage{mathrsfs}%
\usepackage{graphicx}
\usepackage{amsmath}
\usepackage{bm}
\usepackage{feynmf}
\usepackage[citecolor=green]{hyperref}
\def\slash#1{#1 \hskip-0.45em /}
\def\Slash#1{#1 \hskip-0.59em /}

\begin{document}
\title{$\chi_{QJ}\to\ell^{+}\ell^{-}$ within and beyond the Standard Model}

\author{Deshan Yang\inst{1}\thanks{e-mail:yangds@gucas.ac.cn} \and Shuai Zhao\inst{1}
}
\institute{College of Physical Sciences, Graduate University of
Chinese Academy of Sciences, Beijing 100049, China}
\date{Received: date / Revised version: date}
%
\abstract{ We revisit $\chi_{QJ}\to \ell^+\ell^-$ (with $J=0,1,2$
and $Q=b,c$)  within the Standard Model (SM). The electro-magnetic
contributions are given in color-singlet model with non-vanishing
lepton masses at the leading order of $v$. Numerically, the
branching ratios of $\chi_{QJ}\to\ell^{+}\ell^{-}$ predicted within
the SM are so
 small that such decays are barely possible to be detected at future
 BESIII and SuperB experiments, but may be possible to be observed at the LHC.
 We investigate $\chi_{b0}\to\ell^+\ell^-$ in Type-II 2HDM with large $\tan\beta$,
 and $\chi_{b2}\to\ell^+\ell^-$ in the Randall-Sundrum model, to see their chance to be observed in future experiments.
}%
\maketitle

\section{Introduction}

The leptonic decay of quarkonium, especially the S-wave triplet
$J/\psi$ or $\Upsilon$, plays a very important role in particle
physics, either due to its clean experimental signal which is
commonly used to tag $J/\psi$ or $\Upsilon$ in experiments, or due
to its simpleness in theoretical calculations which offers an ideal
place for precise determinations of the non-perturbative
Non-relativistic QCD (NRQCD) matrix elements \cite{Bodwin:1994jh}.
Thus, $H(^3S_1)\to \ell^+\ell^-$ has been extensively studied in
past almost four decades
\cite{NLO,Beneke:1997jm,Czarnecki:1997vz,Brambilla:2006ph,Bodwin:2007ga,Marquard:2009bj}.

However, the leptonic decays of C-even quarkonia are of less
interests, because they are generally suppressed in the SM by both
of the  electromagnetic loop and huge mass of $Z$ boson. The
leptonic decay of $\eta_c$  is investigated by the method of
light-cone wave function in \cite{Yang:2009kq} and NRQCD
factorization at leading order of typical quark velocity $v$ in
\cite{Jia:2009ip}.  The leptonic decay of the C-even and P-wave
quarkonia (i.e. $^3P_J$ quarkonium $\chi_{QJ}$ ($J=0,1,2$)) has been
studied with the color singlet model by K\"{u}hn \emph{et al} about
three decades ago \cite{Kuhn:1979bb}.

In this paper, we will revisit $\chi_{QJ}\to \ell^{+}\ell^{-}$
($J=0,1,2$ and $Q=b,c$) within the SM. We employ the color-singlet
model to calculate the decay amplitudes as K\"{u}hn \emph{et al} did
in \cite{Kuhn:1979bb}.  We get the decay amplitudes with finite
lepton mass $m_{\ell}$, which are relevant for the
helicity-suppressed decay $\chi_{Q0}\to \ell^+\ell^-$, and
$\chi_{bJ}$ or higher $^{3}P_{J}$ charmonium excitations decays to
$\tau^{+}\tau^{-}$. As it should be, our results agree with those
obtained in \cite{Kuhn:1979bb} by setting $m_{\ell}\to 0$.
Furthermore, we calculate the electromagnetic loop by utilizing the
method of regions\cite{beneke:Threthold,Smirnov:expansion}. It
allows us to relate our results to those calculated in the
conventional NRQCD factorization for the $P$-wave quarkonia decays
and productions as in
\cite{Petrelli:1997ge,Bodwin:2007zf,Ma:2002eva}.

Phenomenologically, as expected, the decay widths of $\chi_{QJ}\to
\ell^{+}\ell^{-}$ are highly suppressed in the SM. It is barely
possible to measure such decays even in now-days or near-future
high-luminosity colliders, such as BESIII and SuperB, but may be
possible to be  observed at LHC when the LHC reach its long-term
integrated luminosity around 3000 fb$^{-1}$. Therefore, in an era
longing for the new physics (NP), the smallness of the branching
ratios of such decays in the SM can be a virtue.  Any experimental
discovery of excesses of such decays could be an indication of the
NP. Moreover, the quantum numbers of $\chi_{QJ}$ make the NP effects
in $\chi_{QJ}\to \ell^{+}\ell^{-}$ spin-dependent.  We consider two
scenarios of extensions of the SM which may enhance $\chi_{b0}\to
\ell^+\ell^-$ and $\chi_{b2}\to \ell^+\ell^-$: one is Type-II two
Higgs doublet model (2HDM) with large $\tan\beta$
\cite{arXiv:1106.0034}, another is the Randall-Sundrum (RS) model
for warped extra-dimensions \cite{randall-sundrum,randall-wise}.

This paper is organized as follows. In Sect.\ref{sect:loop},  we
calculate the decay amplitudes for $\chi_{QJ}\to \ell^+\ell^-$
within the SM, compare our results with those obtained in
\cite{Kuhn:1979bb}, and discuss briefly the breakdown and
restoration of the NRQCD factorization for $\chi_{QJ}\to
\ell^+\ell^-$ along the way that Beneke and Vernazza did for  $B\to
\chi_{cJ}K$ in \cite{Beneke:2008pi}.  In Sect. \ref{sect:np}, we
calculate the $\chi_{b0}\to \ell^+\ell^-$ in Type-II 2HDM, and
$\chi_{b2}\to \ell^+\ell^-$ in the RS model.  Sect.\ref{sect:num} is
devoted for the numerical results and some phenomenological
discussions.   Finally, we summarize our work in Sect.
\ref{sect:sum}.

\section{The decay amplitudes for $\chi_{QJ}\to \ell^+\ell^-$ within the SM\label{sect:loop}}

\begin{figure*}
\centerline{
\includegraphics[width=8cm]{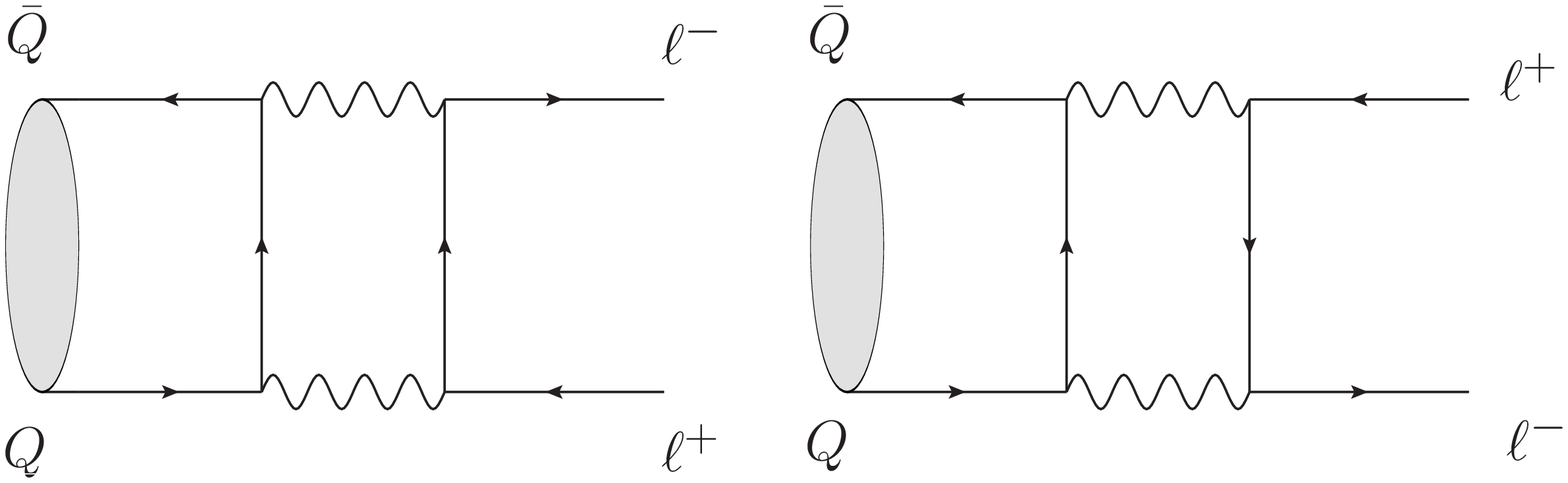}
\includegraphics[width=5cm]{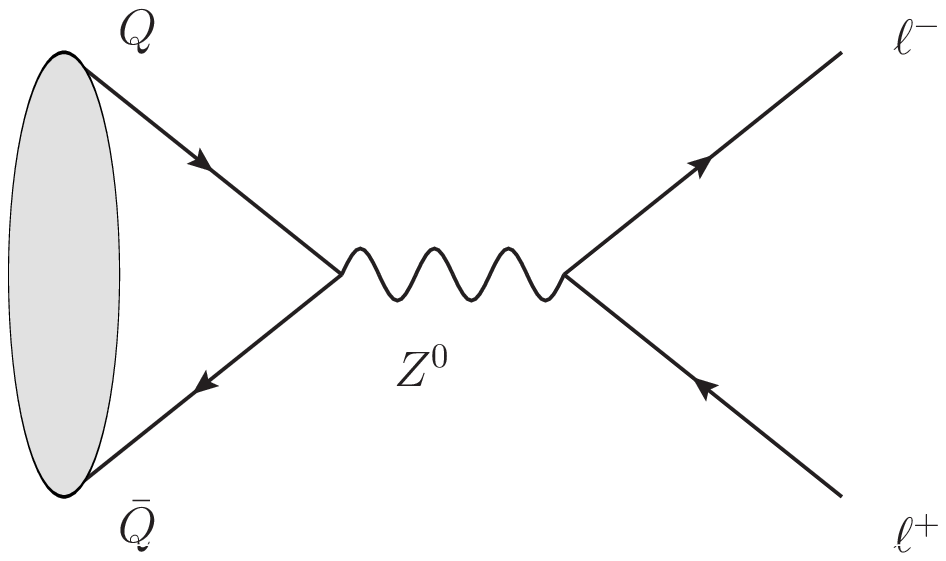}
} \caption{The Feynman diagrams for $^3P_J$ state $Q\bar Q$ direct
annihilation to lepton pair in the SM.  All the Feynman diagrams in
this paper are created by JaxoDraw
\cite{JaxoDraw}.}\label{diag:hard}
\end{figure*}

In the SM, the lowest order Feynman diagrams for $\chi_{QJ}\to
\ell^+\ell^-$ are  two electro-magnetic (EM) box-diagrams and a
tree-level $Z^0$-exchange diagram, depicted in Fig.\ref{diag:hard}.
Note that only the $^3P_1$ state can decay into a lepton pair via
virtual $Z^{0}$ at tree-level, and the tree-level neutral Higgs
exchange diagram for $\chi_{Q0}\to \ell^+\ell^-$ is neglected.

We perform the calculations in the  rest frame of quarkonium.  The
four-momenta of the quark, anti-quark, lepton and anti-lepton are
denoted as
\begin{subequations}
\begin{eqnarray}
\label{momentum} p_1&=&(E,\textbf{q}\,),~~
p_2\,=\,(E,-\textbf{q}\,),\\q_1&=&(E,\textbf{q}_1\,),~~
q_2\,=\,(E,-\textbf{q}_1\,).
\end{eqnarray}
\end{subequations}
where $E=M_{\chi_{QJ}}/2$ with $M_{\chi_{QJ}}$ being mass of
$\chi_{QJ}$. We set $2P$ as the total momentum of $Q\bar Q$, and
$2q$ the relative momentum of $Q\bar Q$, i.e.
\begin{eqnarray}
P=\frac{p_1+p_2}{2} = (E,0)\,,~~
q=\frac{p_1-p_2}{2}=(0,\textbf{q}).\;
\end{eqnarray}
The velocity of the heavy quark is defined as
$v\equiv|\textbf{q}|/m_Q$.

As in \cite{arXiv:0808.1625,Sang:2009jc}, we describe briefly the
procedure to calculate the P-wave quarkonium involved processes
within the color singlet model developed by K\"uhn \emph{et al} in
\cite{Kuhn:1979bb}.

To project the spin-triplet part of the amplitude \\
$Q \bar Q
(^3P_J) \to \ell^ + \ell^-$, we  use the covariant projection
operator of spin-triplets \cite{Bodwin:2002hg}
\begin{eqnarray}
\Pi^{\nu}(\text{spin-triplet})&=&-\frac{1}{2\sqrt{2}E(E+m_Q)}(\Slash{P}+\slash{q}+m_Q)\nonumber\\
&\times&(\Slash{P}+E)\gamma^{\nu}(\Slash{P}-\slash{q}-m_Q)\,,
\end{eqnarray}
to replace the quark-anti-quark spinor bilinear, where $m_Q$ is the
heavy quark mass. We expand the spin-triplet amplitude as series of
the small momentum $q$
\begin{eqnarray}\label{expansion}
i\mathcal{M}^\nu(q)=i\mathcal{M}^\nu(0)+q_{\mu}\frac{\partial}{\partial
q_{\mu}}i\mathcal{M}^\nu(q)|_{q=0}+\cdots.
\end{eqnarray}
Then, the amplitude at  the leading order of $v$ for a $^3P_J$ state
decay to $\ell^+\ell^-$ can be obtained through
\begin{eqnarray}\label{eq:3pj}
i{\cal M}[^3P_J\to\ell^+\ell^-]=\vert
\textbf{q}\vert\Pi_{\mu\nu}(^3P_J)\frac{\partial}{\partial
q_{\mu}}i\mathcal{M}^\nu(q)|_{q=0}\,,
\end{eqnarray}
with the projectors
\begin{subequations}
\begin{eqnarray}
\label{project}
&&\Pi_{\mu\nu}(^3P_0)=\frac{1}{\sqrt{3}}\left(-g_{\mu\nu}+\frac{P_{\mu}P_{\nu}}{E^2}\right),\\
&&\Pi_{\mu\nu}(^3P_1)=\frac{i}{\sqrt{2}E}\epsilon_{\mu\nu\rho\sigma}P^{\rho}\epsilon^{\sigma},\\
&&\Pi_{\mu\nu}(^3P_2)=\epsilon_{\mu\nu}\,,
\end{eqnarray}
\end{subequations}
where $\epsilon_{\sigma}$ and $\epsilon^{\mu\nu}$ are respectively
the polarization vector and tensor for the $^3P_1$ and $^3P_2$
states, which satisfy the relations
$P^{\mu}\epsilon_{\mu}=0,~P^{\mu}\epsilon_{\mu\nu}=0$ and
$\epsilon_\mu^{~\mu}=0$. The polarization summations are
\cite{Kuhn:1979bb}
\begin{subequations}
\begin{align}
&\sum
\epsilon_{\mu}\epsilon_{\nu}=-g_{\mu\nu}+\frac{P_{\mu}P_{\nu}}{E^2}\equiv\mathbb{P}_{\mu\nu}\label{polarization sum:vec}\,,\\
&\sum\epsilon_{\mu\nu}\epsilon_{\alpha\beta}=\frac{1}{2}\left[\mathbb{P}_{\mu\alpha}\mathbb{P}_{\nu\beta}+\mathbb{P}_{\mu\beta}\mathbb{P}_{\nu\alpha}\right]-\frac{1}{3}\mathbb{P}_{\mu\nu}\mathbb{P}_{\alpha\beta}\,.\label{polarization
sum:ten}
\end{align}
\end{subequations}
After convoluting the amplitudes in (\ref{eq:3pj}) with the radial
wave-function for $\chi_{QJ}$ in momentum-space, we reach the final
amplitudes for $\chi_{QJ}\to \ell^+\ell^-$ which is
\begin{eqnarray}\label{eq:Rp0}
i{\cal M}[\chi_{QJ}\to\ell^+\ell^-]&=&-\sqrt{2
M_{\chi_{QJ}}N_c}\sqrt{\frac{3}{4\pi}}R^\prime(0)\nonumber\\
&\times&\Pi_{\mu\nu}(^3P_J)\frac{\partial}{\partial
q_{\mu}}i\mathcal{M}^\nu(q)|_{q=0}\,,
\end{eqnarray}
where the pre-factor $\sqrt{2 M_{\chi_{QJ}}}$  originates from the
relativistic normalization of the $\chi_{QJ}$ state, $\sqrt{N_c}$
from summation of the colors with $N_c=3$ being the number of
colors, and $R^\prime(0)$ is derivative of the radial wave function
of $\chi_{QJ}$  at origin in coordinate-space  \cite{Kuhn:1979bb}.

\subsection{Calculation of the box diagrams}

We calculate the two box-diagrams depicted in Fig. \ref{diag:hard}
by use of the method of regions to
\cite{beneke:Threthold,Smirnov:expansion}. The leading regions in
these box diagrams for the $^3P_J$ state decays into lepton pair in
expansion of the relative velocity $v$, are: 1) hard region where
each component of the momenta of both photons in the loop at order
of $m_Q$; 2) ultra-soft region where each component of the momentum
of one photon at order of $m_Q v^2$. In each region, the power
counting rules of all momenta are clear so that we can perform the
$v$-expansion for the loop-integrands straightforwardly. Then we
integrate the expanded loop-integrand over the whole momentum space
to get the contributions from each region, which sum reproduce the
complete amplitudes in a series of $v$.

With all the techniques described above,  after some tedious but
straightforward calculation, we get the amplitudes from the hard
region which are infrared  (IR) divergent. Here we use the
dimensional regularization (DR) to regulate the IR divergence. The
corresponding amplitudes are
\begin{subequations}\label{eq:hard}
\begin{align}
&\label{hard:0} i\mathcal{M}_{\text{hard}}[^3P_0\to
\ell^+\ell^-]=-i\frac{2\sqrt{6}\alpha^2e^2_Q}{3\beta
m_Q^3}\nonumber\\
&\times\left[\left(\ln\frac{1+\beta}{1-\beta}+\frac{2\beta-\ln\frac{1+\beta}{1-\beta}}{\beta^2}\right)\frac{1}{\epsilon_{\text{IR}}}+\text{finite terms}\right]\nonumber\\
&\times|\textbf{q}|m_{\ell} \bar u(q_1) v(q_2)\,, \end{align}
\begin{align}
&\label{hard:1} i\mathcal{M}_{\text{hard}}[^3P_1\to
\ell^+\ell^-]=-i\frac{2\alpha^2e^2_Q}{\beta m_Q^2}\nonumber\\
&\times\left[\left(\ln\frac{1+\beta}{1-\beta}+\frac{2\beta-\ln\frac{1+\beta}{1-\beta}}{\beta^2}\right)\frac{1}{\epsilon_{\text{IR}}}+\text{finite terms}\right]\nonumber\\
&\times|\textbf{q}|\bar u(q_1)\gamma_5\slash\epsilon
v(q_2)\,,\end{align}
\begin{align}&\label{hard:2}
i\mathcal{M}_{\text{hard}}[^3P_2\to
\ell^+\ell^-]=i\frac{2\sqrt{2}\alpha^2e^2_Q}{\beta
m_Q^3}\nonumber\\
&\times\left[\left(\ln\frac{1+\beta}{1-\beta}+\frac{2\beta-\ln\frac{1+\beta}{1-\beta}}{\beta^2}\right)\frac{1}{\epsilon_{\text{IR}}}+\text{finite terms}\right]\nonumber\\
&\times|\textbf{q}|\bar
u(q_1)\gamma^{\alpha}v(q_2)q^{1\beta}\epsilon_{\alpha\beta}\,,
\end{align}
\end{subequations}
where $\beta\equiv\sqrt{1-m^2_{\ell}/m^2_Q}$ is the velocity of
lepton, $\alpha=e^2/(4\pi)$ the fine-structure constant, and $e_Q$
the electric charge in unit of elementary charge $e$ for the heavy
quark $Q$. We do not list the explicit expressions for the finite
terms above, since they are somewhat scheme-dependent and therefore
meaningless individually.  One should notice that
$i\mathcal{M}_{\rm hard}[^3P_0\to\ell^+\ell^-]$ is proportional to
$m_\ell$ from the helicity suppression.

In a similar way, we get the contributions from the ultra-soft
region to the amplitudes which are ultraviolet (UV) divergent. Here
we also use the (DR) to regulate the UV divergence. The
corresponding amplitudes are
\begin{subequations}\label{eq:soft}
\begin{align}
&\label{soft:0} i\mathcal{M}_{\text{us}}[^3P_0\to
\ell^+\ell^-]=i\frac{2\sqrt{6}\alpha^2e^2_Q}{3\beta
m_Q^3}\nonumber\\
&\times\left[\left(\ln\frac{1+\beta}{1-\beta}+\frac{2\beta-\ln\frac{1+\beta}{1-\beta}}{\beta^2}\right)\frac{1}{\epsilon_{\text{UV}}}+\text{finite terms}\right]\nonumber\\
&\times|\textbf{q}|m_{\ell} \bar u(q_1) v(q_2)\,,
\end{align}
\begin{align}
&\label{soft:1}i\mathcal{M}_{\text{us}}[^3P_1\to
\ell^+\ell^-]=i\frac{2\alpha^2e^2_Q}{\beta m_Q^2}\nonumber\\
&\times \left[\left(\ln\frac{1+\beta}{1-\beta}
+\frac{2\beta-\ln\frac{1+\beta}{1-\beta}}{\beta^2}\right)
\frac{1}{\epsilon_{\text{UV}}}+\text{finite terms}\right]\nonumber\\
&\times|\textbf{q}|\bar u(q_1)\gamma_5\slash\epsilon
v(q_2)\,,
\end{align}
\begin{align}
&\label{soft:2} i\mathcal{M}_{\text{us}}[^3P_2\to
\ell^+\ell^-]=-i\frac{2\sqrt{2}\alpha^2e^2_Q}{\beta m_Q^3}\nonumber\\
&\times
\left[\left(\ln\frac{1+\beta}{1-\beta}+\frac{2\beta-\ln\frac{1+\beta}{1-\beta}}{\beta^2}\right)\frac{1}{\epsilon_{\text{UV}}}+\text{finite terms}\right]\nonumber\\
&\times|\textbf{q}|\bar
u(q_1)\gamma^{\alpha}v(q_2)q^{1\beta}\epsilon_{\alpha\beta}\,.
\end{align}
\end{subequations}
It is easy to see from (\ref{eq:hard}) and (\ref{eq:soft}), that the
divergent parts of hard and ultra-soft parts have opposite signs.
Thus, the whole amplitudes for $^3P_J\to \ell^+\ell^-$ due to the EM
interactions
\begin{eqnarray}
i\mathcal{M}_{em}[^3P_J\to
\ell^+\ell^-]&=&i\mathcal{M}_\text{hard}[^3P_J\to
\ell^+\ell^-]\nonumber\\
&+&i\mathcal{M}_\text{us}[^3P_J\to \ell^+\ell^-]
\end{eqnarray}
are finite.

 In all, we have
\begin{eqnarray}\label{eq:amp}
i\mathcal{M}_{em}[\chi_{QJ}\to \ell^+ \ell^-]&=& -i
{\sqrt{2M_{\chi_{QJ}} N_c}}\sqrt{\frac{3}{4\pi}}R'(0)\frac{e^2_Q
\alpha^2}{m_Q^4}f_{J}\nonumber\\
&\times&\left\{\begin{array}{ll}
 m_\ell\bar
u(q_1) v(q_2)\,, & J=0\,,\\
m_Q\bar u(q_1)\gamma_5\slash\epsilon v(q_2)\,,& J=1\,,\\\bar
u(q_1)\gamma^{\alpha}v(q_2)q^{1\beta}\epsilon_{\alpha\beta}\,, &
J=2\,.
\end{array}\right.\nonumber\\
\end{eqnarray}
The finite coefficients $f_J(J=0,1,2)$ are from the sums of
(\ref{eq:hard}) and (\ref{eq:soft}), their explicit analytic
expressions are
\begin{subequations}
\begin{align}
f_0&=\frac{2 \sqrt{6} \left(2 \beta-\left(1-\beta ^2\right) \ln
\left(\frac{\beta +1}{1-\beta }\right) \right) \ln
\left(\frac{m_Q}{\omega}\right)}{3\beta
^3}\nonumber\\
&+\frac{\sqrt{6}
   \left(\left(\beta ^2+2\right) \text{Li}_2\left(\frac{\beta -1}{\beta +1}\right)-2 \left(\beta ^2-1\right) \text{Li}_2\left(\frac{1-\beta }{\beta +1}\right)\right)}{3\beta ^3}\nonumber\\
   &+\frac{1}{2\sqrt{6}\beta^3}(4-\beta^2)\ln^2\left(\frac{\beta+1}{1-\beta}\right)\nonumber\\
   &+\frac{1}{6\sqrt{6}\beta^3}\left[\pi^2(5\beta^2-2)+12\beta\ln(1-\beta^2)+24\beta\right]\nonumber\\
   &+\frac{1}{\sqrt{6}\beta^3}\ln\left(\frac{\beta+1}{1-\beta}\right)\left[(-2-3i\pi)\beta^2-4(1-\beta^2)\ln(4\beta)\right.\nonumber\\
   &\left.+2(1-\beta^2)\ln(1-\beta^2)+2\right]
   ,\label{a1}\\
f_1&=\frac{2 \left(2 \beta-\left(1-\beta ^2\right) \ln
\left(\frac{\beta +1}{1-\beta }\right) \right) }{\beta ^3}\ln
\left(\frac{m_Q}{\omega}\right)\nonumber\\
&+\frac{ \left(1-\beta ^2\right)
  }{\beta ^3}\text{Li}_2 \left(\left(\frac{1-\beta }{\beta
  +1}\right)^2\right)\nonumber
\end{align}
\begin{align}
   &-\frac{2 \left(1-\beta ^2\right) }{\beta ^3}\left[\left(\ln
   \left(\frac{4 \beta }{\beta +1}\right)-1\right) \ln \left(\frac{\beta +1}{1-\beta }\right)+\frac{\pi
   ^2}{12}\right],\label{a2}\\
f_2&=\frac{2\sqrt{2}}{\beta^3}\left(2\beta-(1-\beta^2)\ln \left(
\frac{\beta+1}{1-\beta}\right)\right)
   \ln \left(\frac{m_Q}{\omega }\right)\nonumber\\
   &-\frac{\sqrt{2} \left(\beta ^2-1\right)^2
\text{Li}_2\left(\frac{\beta -1}{\beta +1}\right)}{\beta ^5}-\frac{2
\sqrt{2} \left(\beta ^2-1\right) \text{Li}_2\left(\frac{1-\beta
}{\beta
   +1}\right)}{\beta ^3}\nonumber\\
   &-\frac{\left(1-3 \beta ^2\right) \left(1-\beta ^2\right) \ln
^2\left(\frac{1-\beta }{\beta +1}\right)}{2 \sqrt{2} \beta
^5}+\frac{1}{\sqrt{2}\beta^5}\ln \left(\frac{1-\beta }{\beta
+1}\right)\nonumber\\
&\times \left[2\beta^2(1-\beta^2)\left(4\ln
2-1-\ln\left(\frac{1-\beta^2}{\beta^2}\right)\right)\right.\nonumber\\
&\left.+(\beta^4-1)i\pi+4\beta^4
  \right]+\frac{1}{6\sqrt{2}\beta^5}\left[-\pi^2(1-\beta^2)(1+3\beta^2)\right.\nonumber\\
  &\left.+4\beta(3+\beta^2)(2\ln 2-i\pi)\right.\nonumber\\
   &\left.+16\beta^3-12\beta(1+\beta^2)\ln(1-\beta^2)\right]\,.
   \label{a3}
\end{align}
\end{subequations}
Here, $\omega=M_{\chi_{QJ}}-2m_Q+i\epsilon$ is the binding energy of
$\chi_{QJ}$, and $\text{Li}_2(x)=-\int_0^x dt\ln(1-t)/t$ is the
dilogarithm function.

In the massless lepton limit $r\equiv m_\ell^2/m_Q^2={1-\beta^2}\to
0$,  \begin{subequations}
\begin{align}
&f_0\to
\frac{1}{2\sqrt{6}}\left[16\ln\left(\frac{m_Q}{\omega}\right)+3\ln^2
r+\left(4+6
i\pi-12\ln 2\right)\ln r\right.\nonumber\\
&\left.~~~~~~+\pi^2+8+12\ln^2 2-12 i \pi \ln 2\right],\\
&f_1\to 4\ln\frac{m_Q}{\omega},\\
&f_2\to-\frac{4\sqrt{2}}{3}\left(\ln{2}-3\ln\frac{m_Q}{\omega}-1+i\pi
\right)\,.
\end{align}
\end{subequations}
 One can see that $f_0$ diverges when $r\to 0$ while $f_{1,2}$ remain finite in the same limit. However, the decay amplitude for $\chi_{Q0}\to \ell^+\ell^-$ still vanishes when $r\to 0$, since we have singled out the helicity-suppression factor $m_\ell$.  Finally, we reproduce the
results of K\"uhn \emph{et al} in \cite{Kuhn:1979bb}:
\begin{subequations}\label{eq:0mass}
\begin{align}
i\mathcal{M}_{em}[\chi_{Q0}&\to \ell^+ \ell^-]\to 0, \\
i\mathcal{M}_{em}[\chi_{Q1}&\to \ell^+
\ell^-]\to-i\sqrt{2M_{\chi_{Q1}}N_c}\sqrt{\frac{3}{4\pi}}R'(0)\nonumber\\
&\times\frac{4e^2_Q \alpha^2}{m^3_Q}\ln\frac{m_Q}{\omega}\bar
u(q_1)\gamma_5\slash\epsilon
v(q_2),\\
i\mathcal{M}_{em}[\chi_{Q2}&\to \ell^+
\ell^-]\to-i\sqrt{2M_{\chi_{Q1}}N_c}\sqrt{\frac{3}{4\pi}}R'(0)\nonumber\\
&\times\frac{4\sqrt{2}e^2_Q \alpha^2}{m^4_Q}\left(\ln\frac{m_Q}{\omega}+\frac{1-\ln 2-i\pi}{3}\right)\nonumber\\
&\times\epsilon_{\mu\nu}\bar u(q_1)\gamma^{\mu}
v(q_2)\frac{(q_1-q_2)^{\nu}}{2}.
\end{align}
\end{subequations}

\begin{figure}[h]
\centerline{\includegraphics[width=9cm]{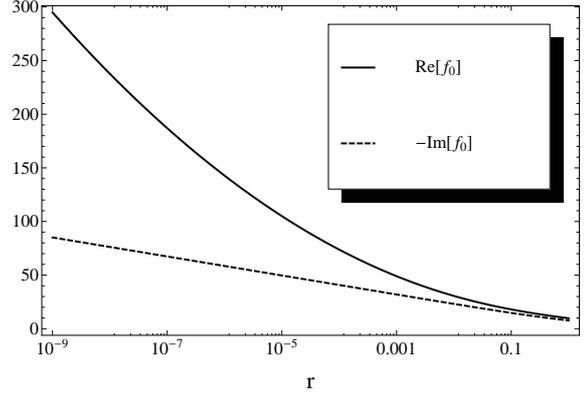}}\centerline{\includegraphics[width=9cm]{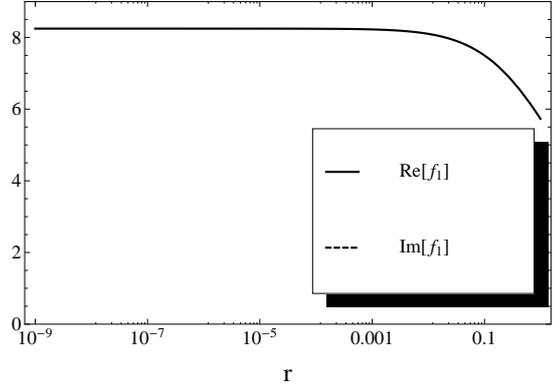}}\centerline{\includegraphics[width=9cm]{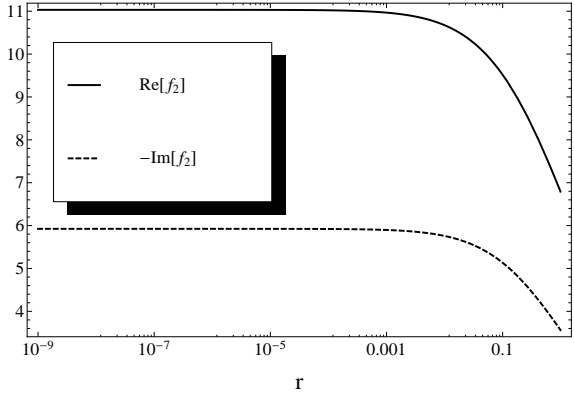}}
\caption{ The real and imaginary parts of $f_i$ as functions of
$r=m_\ell^2/m_c^2$. }\label{chij}
\end{figure}

Taking $\chi_{cJ}$ as examples, we plot  the real and imaginary
parts of the coefficients $f_J$ ($J=0,1,2$) as functions of
$r=m_{\ell}^2/m_c^2$ in Fig. \ref{chij}.  We roughly take the
$m_c=1.65$ GeV and $M_{\chi_{c1,2,3}}=3.41, 3.51, 3.56$ GeV.
$f_{1,2}$ are almost flat when $r\leq 10^{-2}$, but decrease
evidently where $r$ is greater than several percent. Thus, our
results are useful to make more accurate predictions for $\chi_{bJ}$
or higher radial excitations of $\chi_{cJ}$ decays to
$\tau^+\tau^-$.

Moreover, theoretically, the imaginary parts of $f_{J}$ originate
from the on-shell immediate states. In Fig. \ref{chij}, we see that
the imaginary part of $f_{1}$ vanishes because a massive
(axial)-vector cannot decay to two photons due to Yang's theorem
\cite{c.n.yang}. However, one should notice that, if the binding
energy $\omega$ can be negative,  the $\ln {m_Q/\omega}$ term in
$f_1$ can contribute an imaginary part. And the $\ln \omega$ term
comes only from the ultra-soft part in (\ref{eq:soft}).
 As K\"{u}hn \emph{et al} have pointed out in \cite{Kuhn:1979bb}, that this effect is related to the E1 transition $^3P_J\to ^3S_1+\gamma$.

\subsection{The neutral current contributions\label{sect:EW}}
For completeness, we present the neutral current weak interaction
contributions to  $\chi_{Q1}\to \ell^+\ell^-$ as follows.
\begin{align}
i\mathcal{M}_{weak}[\chi_{Q1}&\to \ell^+\ell^-]=\pm i  G_F
\sqrt{\frac{3 N_c}{\pi M_{\chi_{Q1}}}}R'(0)\nonumber\\
&\times\bar u(q_1)\slash
\epsilon\left(1-4\sin^2\theta_W-\gamma_5\right)v(q_2)\,,
\end{align}
where the pre-factor ``$+$" is for $Q=c$, ``$-$" for $Q=b$,  $G_F$
the Fermi constant, and $\theta_W$ the Weinberg angle. As we will
see in the numerical analysis,  the neutral-current contribution may
play an important role, especially in  $\chi_{bJ}\to \ell^+\ell^-$
since its electromagnetic decay amplitude is further suppressed by
$e_b^2=1/9$.

In all, the decay widths for $\chi_{QJ}\to\ell^+\ell^-$ within the
SM can be written as
\begin{align}\label{eq:dwsm}
&\Gamma[\chi_{Q0}\to \ell^+\ell^-]=\frac{3N_c e^4_Q \alpha^4}{4\pi^2
m^4_Q}\beta^3(1-\beta^2)|f_0|^2|R'(0)|^2\,,\\
&\Gamma[\chi_{Q1}\to \ell^+\ell^-]=\frac{\beta N_c}{8\pi^2}|R '(0)|^2\left[\frac{4e^4_Q \alpha^4}{m^4_Q}|f_1|^2\beta^2\right.\nonumber\\
&\left.+ G_F^2\left(2\beta^2 g^2_a+(3-\beta^2)g^2_v\right)
\pm\frac{4\sqrt{2} e^2_Q \alpha^2G_F}{m_Q^2
}\beta^2g_a \text{Re}[f_1]\right]\,,\\
&\Gamma[\chi_{Q2}\to \ell^+\ell^-]=\frac{N_c e^4_Q \alpha^4}{20\pi^2
m^4_Q}\beta^3(5-2\beta^2)|f_2|^2|R'(0)|^2\,,
\end{align}
where $g_v={1-4\sin^2\theta_W}$, $g_a=-1$, and ``$\pm$" correspond
to $Q=c$ and $Q=b$, respectively.

\subsection{Connections to the NRQCD factorization\label{sect:lagrangian}}

Before we get into phenomenological applications of our results
obtained above, we would like to translate our calculations into the
language of the effective field theory, and see what we can get from
such comparisons.

Heavy quarkonium decays involve three well-separated intrinsic
scales $m_{Q},~m_{Q}v,~m_{Q}v^2$. The NRQCD is a suitable and
powerful effective field theory to describe the heavy quarkonia
production and decays \cite{Bodwin:1994jh}.
 By integrating out the hard fluctuations around the scale $m_{Q}$, the effective
Lagrangian for the leptonic decay of a non - relativistically moving
heavy quark pair $Q\bar Q$ can
 be written as
\begin{align}
\label{4-fermion}
\delta\mathcal{L}=&\frac{f(^3S_1)}{m^2_{Q}}\mathcal{O}(^3S_1)
+\sum\limits_{J=0,1,2}\frac{f(^3P_J)}{m^4_{Q}}\mathcal{O}(^3P_J)
+\cdots\,,
\end{align}
where $f(^{2S+1}L_{J})$ are the short-distance coefficients supposed
to be finite, and the effective operators are
\begin{subequations}
\begin{align}
\label{operator}
&\mathcal{O}(^3S_1)=\chi^\dag\gamma^{i}\psi\bar \ell\gamma_{i}\ell,\\
&\mathcal{O}(^3P_0)=-\frac{2
m_\ell}{\sqrt{3}}\chi^\dag\left(-\frac{i}{2}\stackrel{\leftrightarrow}{\Slash{D}}\right)\psi\bar
\ell\ell\, \\
&\mathcal{O}(^3P_1)=-\frac{1}{\sqrt{2}}\chi^\dag\left(-\frac{i}{2}\stackrel{\leftrightarrow}{D}\!{}^i\left[\gamma_i,\gamma_j\right]\gamma_5\right)
\psi\bar \ell\left[i\not\! \stackrel{\leftrightarrow}{\partial},\gamma^{j}\right]\ell,\\
&\mathcal{O}(^3P_2)=\chi^\dag\left[\left(-\frac{i}{2}\right)\stackrel{\leftrightarrow}{D}
\!{}^{(i}\gamma^{j)}\right]\psi\bar \ell\, i
\stackrel{\leftrightarrow}{\partial}_{(i}\gamma_{j)}\ell.
\end{align}
\end{subequations}
Here, following the conventions adopted in \cite{Beneke:2008pi}, we
employ the four-component spinors fields $\psi$ and $\chi$ to
represent the the non-relativistic heavy quark and anti-quark,
respectively. These fields satisfy $\gamma^0 \psi=\psi$ and
$\gamma^0\chi=-\chi$, and are equivalent to the conventional NRQCD
two-component fields used in \cite{Bodwin:1994jh}. The Latin indices
$i,j$ run over the spacial indices $1,2,3$, and $(i,j)$ means the
traceless part of a symmetric tensor.  And
\begin{eqnarray}
\chi^\dag\stackrel{\leftrightarrow}{D}\!{}^i\Gamma\psi&\equiv&\chi ^\dag D^{i}\Gamma\psi-\chi^\dag\stackrel{\leftarrow}D\!{}^{i}\Gamma\psi\,,\\
\bar \ell
\stackrel{\leftrightarrow}{\partial}\!{}^\mu\Gamma\ell&\equiv& \bar
\ell {\partial}\!{}^\mu\Gamma\ell-\bar \ell
\stackrel{\leftarrow}{\partial}\!{}^\mu\Gamma\ell\,,
\end{eqnarray}
where $\Gamma$ denotes any Dirac structure,
$D^{i}\equiv\partial^{i}-ig_{s}A^{a,i}T^{a}$ the covariant
derivative in QCD. Note that the $P$-wave operators ${\cal
O}(^{3}P_{J})$ are $v$-suppressed relative to the $S$-wave operator
${\cal O}(^{3}S_{1})$ due to the power-counting rules in NRQCD.

Neglecting the contributions from weak interaction,  at the lowest
order of the strong coupling $\alpha_{s}$, we have the expectation
$f(^{3}S_{1})$ at $O(\alpha)$ while $f(^3P_J)$ at $O(\alpha^2)$.
Naively, one would expect that
\begin{eqnarray}\label{eq:nrfact}
&&i\mathcal{M}[\chi_{QJ}\to
\ell^+\ell^-]=i\frac{f(^3P_J)}{m^4_{Q}}\langle\ell^+\ell^-\vert\mathcal{O}(^3P_J)\vert
\chi_{QJ}\rangle\,,\nonumber\\
\end{eqnarray}
within the frame of NRQCD at the leading order of $v$. Then, the
validation of  (\ref{eq:nrfact}) implies a factorization, in which
the short-distance contributions are absorbed into $f(^3P_J)$ while
all the long-distance contributions are absorbed into the
matrix-elements of $\mathcal{O}(^3P_J)$.

However, one should notice that, within the NRQCD, an ultra-soft
photon can interact with the heavy quarks, and such interaction can
be described by \cite{Pineda:1997bj,Brambilla:1999xf}
\begin{eqnarray}\label{eq:E1}
\delta
\mathcal{L}_{us}&=&ee_Q\psi^\dag(x)\left[A^0_{\text{em}}(t)-\mathbf{x}\cdot
\mathbf{E}(t)\right]\psi(x)\nonumber\\
&+&ee_Q\chi^\dag(x)\left[A^0_{\text{em}}(t)-\mathbf{x}\cdot
\mathbf{E}(t)\right]\chi(x)\,,
\end{eqnarray}
where $A_{\rm em}^0$ is the electro-potential, and $\mathbf{E}$ the
electro-field strength. Both of $A_{\rm em}^0$ and $\mathbf{E}$ have
been multipole expanded. In real calculations of the amplitudes, the
$A^0$ term in (\ref{eq:E1}) can be dropped out either by choosing
the temporal gauge $A^0_{\rm em}=0$ or by the automatic
cancellations in the amplitudes in other gauge choice. The
$\mathbf{E}$ term in (\ref{eq:E1}) is actually the electro-dipole
interaction,  which is at the order $O(v)$ and conserves the spin
but change the orbit angular momentum $L$ by one unit. It implies
that the ultra-soft photon interaction in (\ref{eq:E1}) can
transform a $^3S_1$ state into a $^3P_J$ state with a price of
$O(v)$ suppression.

\begin{figure}[!htb]
\centerline{
\includegraphics[width=8cm]{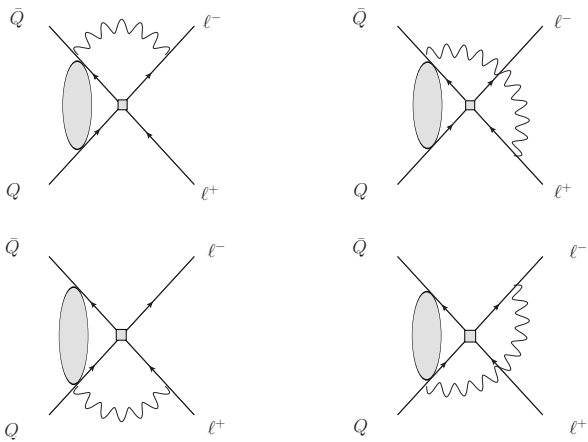}
} \caption{The ultra-soft loop corrections to $\langle
\mathcal{O}(^3S_1)\rangle$. } \label{diag:soft}
\end{figure}

Therefore, the correct decay amplitude for $\chi_{QJ}\to
\ell^+\ell^-$ at the lowest order of $v$ within  the NRQCD should be
written as
\begin{eqnarray}\label{eq:amp}
i\mathcal{M}[\chi_{QJ}\to \ell^+\ell^-]&=&i\frac{f(^3P_J)}{m^4_{Q}}\langle\ell^+\ell^-\vert\mathcal{O}(^3P_J)\vert \chi_{QJ}\rangle|_{\text{tree}}\nonumber\\
&+&i\frac{f(^3S_1)}{m^2_{Q}}\left.\langle\ell^+\ell^-\vert\mathcal{O}(^3S_1)\vert\chi_{QJ}\rangle\right\vert_{\text{us-loop}}.\nonumber\\
\end{eqnarray}
The matrix-element of $\mathcal{O}(^3S_1)$ above can be depicted by
the NRQCD Feynman diagrams in Fig. \ref{diag:soft}.

One familiar with the method of regions, can immediately recognize
the matching equations
\begin{subequations}\label{eq:region}
\begin{align}
i\frac{f(^3P_J)}{m^4_{Q}}\langle\ell^+\ell^-\vert\mathcal{O}(^3P_J)\vert
\chi_{QJ}\rangle|_{\text{tree}}=i\mathcal{M}_{\rm hard}[\chi_{QJ}\to
\ell^+\ell^-]\,,\\
i\frac{f(^3S_1)}{m^2_{Q}}\left.\langle\ell^+\ell^-\vert\mathcal{O}(^3S_1)\vert\chi_{QJ}\rangle\right\vert_{\text{us-loop}}=i\mathcal{M}_{\rm
us}[\chi_{QJ}\to \ell^+\ell^-].\nonumber\\
\end{align}
\end{subequations}
Consequently, one can find that the `` short-distance"  $f(^3P_J)$
 does contain IR divergences which breaks down the naive
factorization in (\ref{eq:nrfact}).

Of course,  the breakdown of the conventional NRQCD factorization
for many $P$-wave quarkonium involved processes is not new. Taking
$B\to \chi_{cJ}K$ for instance, the QCD factorization breaks down
\cite{Song:2002mh,Song:2003yc,Pham:2005ih,Meng:2005fc,Meng:2005er,Meng:2006mi}.
However, Beneke and Vernazza showed in \cite{Beneke:2008pi}, that
the factorization for $B\to \chi_{cJ}K$ can be restored, by
considering the contribution from $S$-wave color-octet operators in
which the chromo-E1 transition analogue to (\ref{eq:E1}) plays a
crucial role.

Therefore, the identifications in (\ref{eq:region}) and finiteness
of (\ref{eq:amp}) can be regarded as an application of the idea
developed by Beneke and Vernazza, to restore the factorization.
However, since it deviates from our final phenomenological goal, we
would like to stop the further discussions along this line.

\section{Possible impacts from new physics beyond the SM\label{sect:np}}

As we have seen, $\chi_{QJ}\to \ell^{+}\ell^{-}$ is highly
suppressed in the SM, due to either the EM loop or the large mass of
$Z^0$. For $\chi_{Q0}\to \ell^{+}\ell^{-}$, it suffers more
suppressions in the SM due to the helicity selection rules. The
tininess of the branching ratios make such decays sensitive to the
possible new physics beyond the SM. If the quantum numbers of the
new particles in the SM extensions match those of $\chi_{QJ}$, and
the couplings among them are enhanced in some way, we may have a
chance to find the hints of new physics  in $\chi_{QJ}\to
\ell^{+}\ell^{-}$. In this section, we consider two kinds of models:
Type-II 2HDM with large $\tan \beta$  \cite{arXiv:1106.0034},  and
the RS model of the warped extra-dimension
\cite{randall-sundrum,randall-wise}.

\subsection{$\chi_{Q0}\to\ell^+\ell^-$ in Type-II 2HDM
\label{chiq0:2HDM}} Type-II 2HDM is one of the most studied
extensions of the SM. It shares almost the same Higgs sector
interactions with the minimal super-symmetric SM (MSSM). The general
Yukawa couplings among the fermions and the lightest neutral Higgs
$h$ in 2HDM can be written as
\begin{align}
{\cal
L}&_Y^{2HDM}=-\frac{g}{2m_W}h\nonumber\\
&\times\left(m_U\frac{\cos\alpha}{\sin\beta}\bar U U
-m_D\frac{\sin\alpha}{\cos\beta}\bar D D
-m_\ell\frac{\sin\alpha}{\cos\beta}\bar\ell \ell\right)\,,
\end{align}
where $U$ denotes for the up-type quarks, $D$ for the down-type
quarks, $\ell$ for the charged leptons, $g$ for the SU(2)$_L$ gauge
coupling, $\alpha$ for the mixing-angle of the neutral Higgs, and
$\tan \beta=v_u/v_d$ with $v_{d,u}$ being the vacuum expectation
values of two Higgs doublets coupled to the down-type and up-type
quarks respectively.

Straightforwardly, we have the amplitudes for $\chi_{Q0}\to
\ell^+\ell^+$ via $h$ in Type-II 2HDM are
\begin{align}
i \mathcal{M}_{2HDM}[\chi_{c0}\to
\ell^+\ell^-]&=i4\sqrt{\frac{6N_c}{m_{\chi_{c0}}}}G_F\frac{\sin
2\alpha~ m_c m_\ell}{\sin 2\beta m^2_h}\nonumber\\
&\times\bar u(q_1) v(q_2)\left(-\sqrt{\frac{3}{4\pi}}R'(0)\right)\,,
\\
i \mathcal{M}_{2HDM}[\chi_{b0}\to
\ell^+\ell^-]&=-i4\sqrt{\frac{6N_c}{m_{\chi_{b0}}}}G_F\frac{\sin
^2\alpha~ m_b m_\ell}{\cos^2 \beta m^2_h }\nonumber\\
&\times\bar u(q_1) v(q_2)\left(-\sqrt{\frac{3}{4\pi}}R'(0)\right)\,,
\end{align}
where $m_h$ is the mass of the lightest neutral Higgs.  Here we have
used the relation $4 \sqrt{2} G_F=g^2/ m_W^2$.

In the large $\tan \beta$ scenario of Type II 2HDM, i.e.
$\tan\beta\gg1$ and $\alpha\sim \beta$, the amplitude for
$\chi_{b0}\to\ell^+\ell^-$ is enhanced by the factor $m_b m_\ell
\tan^2\beta$, which may compensate the suppression from the factor
$m_h^2$, while $\chi_{c0}\to\ell^+\ell^-$ does not receive such
enhancement. Thus, in the numerical analysis below, we will consider
only $\chi_{b0}\to \ell^+\ell^-$ in Type-II 2HDM with large
$\tan\beta$.

\subsection{$\chi_{Q2}\to\ell^+\ell^-$ in the RS model\label{chiq2:RS}}

As a potential
 solution to the hierarchy problem, the Randall - Sundrum(RS) model \cite{randall-sundrum} predicts that a TeV Kaluza-Klein (KK) resonances may couple to
the SM particles. The corresponding effective Lagrangian is
\cite{randall-wise}
\begin{align}
\mathcal{L}_{int}=-\frac{\kappa}{m_0}
h^{KK}_{\mu\nu}T^{\mu\nu}_{SM},
\end{align}
where $h^{KK}_{\mu\nu}$ is the KK graviton field, $T_{SM}$ the SM
energy-momentum stress tensor, $\kappa\equiv k/\overline{M}_{Pl}$
the effective coupling constant with $\overline{M}_{Pl}\equiv
M_{Pl}/\sqrt{8\pi}$ the reduced Plank scale, $k\sim M_{Pl}$ is the
space-time curvature in the extra dimension. $m_0=k e^{-k \pi r_c}$
is a mass scale at the order of TeV, and the KK graviton masses are
$m_i=m_0 x_i$, where $r_c$ is the compatification radius of the
extra dimension, and $x_i$ are roots of Bessel function $J_1(x)$.

The stress tensor for fermions is proportional to \\$\bar
\psi\left[\left(-\frac{i}{2}\right)\stackrel{\leftrightarrow}{D}
{}^{(\mu}\gamma^{\nu)}\right]\psi$, which matches to the quantum
number of a
 $2^{++}$ quarkonium. Thus, $\chi_{Q2}\to\ell^+\ell^-$ can happen at tree-level. Here we consider the contributions from the lowest KK excitations of graviton. A straightforward calculation shows that
\begin{align}\label{eq:rsamp}
i \mathcal{M}_{RS}[\chi_{Q2}&\to \ell^+\ell^-]=i 2\sqrt{2}
\frac{\kappa^2}{m^2_1 m^2_0} \bar u(q_1)\gamma^{\alpha}
v(q_2)\nonumber\\
&\times\frac{(q_1-q_2)_{\beta}}{2}\epsilon_{\alpha\beta}\times\left(-\sqrt{\frac{3}{4\pi}}R'(0)\right),
\end{align}
and the resulted decay width is
\begin{align}\label{eq:rswidth}
\Gamma^{RS}=\frac{32\beta^3
N_c}{5\pi^2}|R'(0)|^2\left(\frac{\kappa}{m_1}\right)^4
\frac{m_Q^4}{m_0^4}(5-2\beta^2),
\end{align}
where $m_1$ is the mass of the lightest KK graviton. Though the KK
graviton contribution is TeV scale suppressed, nevertheless,
$\chi_{Q2}\to\ell^+\ell^-$ may be sizable if $\kappa$ is not too
small. Since  (\ref{eq:rswidth}) indicates that $\Gamma^{RS}$ is
proportional to $m^4_Q$, we only consider $\chi_{b2}\to
\ell^+\ell^-$ in numerical analysis below.

\section{Numerical results and discussions\label{sect:num}}

Here we present the numerical results  based on our calculations of
$\chi_{QJ}\to\ell^+\ell^-$ within the SM and beyond.  We take the
following values for the fine structure constant, Fermi constant and
Weinberg angle\begin{eqnarray} &&\alpha=1/137\,,~~G_F=1.16\times
10^{-5}~\text{GeV}\,,~~\nonumber\\&&
\sin^2\theta_W=0.231\,,
\end{eqnarray}
in our numerical analysis.

\subsection{$\chi_{QJ}\to \ell^+\ell^-$ within the SM}

\subsubsection{$\chi_{cJ}\to \ell^+\ell^-$}

For $\chi_{cJ}$, we take
\begin{eqnarray} &&M(\chi_{c0})=3.41~\text{GeV}\,,~~
M(\chi_{c1})=3.51~\text{GeV}\,,~~\nonumber\\
&&M(\chi_{c2})=3.56~\text{GeV}\,,\\
&&\Gamma(\chi_{c0})=10.3~\text{MeV}\,,~~
\Gamma(\chi_{c1})=0.86~\text{MeV}\,,~~\nonumber\\
&&\Gamma(\chi_{c2})=1.97~\text{MeV}\,.
\end{eqnarray}
 The binding energies are taken as
$\omega =M(\chi_{cJ})-2m_c$.  The derivative of $\chi_{cJ}$ radial
wave function at the origin is taken as
$|R_{\chi_c}'(0)|^2=0.050$GeV$^5$ \cite{Beneke:2008pi}.

In Table \ref{Table:chic:r}, we show the branching ratios of
$\chi_{cJ}\to e^+e^-$ and $\mu^+\mu^-$ with $m_c=1.65$GeV.  One can
see little difference between the $\text{Br}[\chi_{c1,2}\to e^+e^-]$
and $\text{Br}[\chi_{c1,2}\to\mu^+\mu^-]$, but large differences
between $\text{Br}[\chi_{c0}\to e^+e^-]$ and
$\text{Br}[\chi_{c0}\to\mu^+\mu^-]$, because $\chi_{c0}\to
\ell^+\ell^-$ is helicity supressed.

\begin{table}[h]
\caption{\label{Table:chic:r}%
The branching ratios of $\chi_{cJ}\to \ell^+\ell^-$.}
\begin{center}
\begin{tabular}{lcc}
\hline\noalign{\smallskip}
Decay Channels  & QED  & QED+weak \\
\hline
$\chi_{c0}\to e^+e^-$    & $3.19\times10^{-13}$ &\\
$\chi_{c0}\to\mu^+\mu^-$  & $7.06\times10^{-10}$ &\\
\hline
$\chi_{c1}\to e^+e^-$    & $4.55\times10^{-8}$ &$3.67\times10^{-8}$\\
$\chi_{c1}\to\mu^+\mu^-$   & $4.44\times10^{-8}$ &$3.60\times10^{-8}$\\
\hline
$\chi_{c2}\to e^+e^-$    & $1.37\times10^{-8}$ &\\
$\chi_{c2}\to\mu^+\mu^-$   & $1.32\times10^{-8}$ &\\
\noalign{\smallskip}\hline
\end{tabular}
\end{center}
\end{table}

\begin{figure}[!htb]
\centerline{\includegraphics[width=9cm]{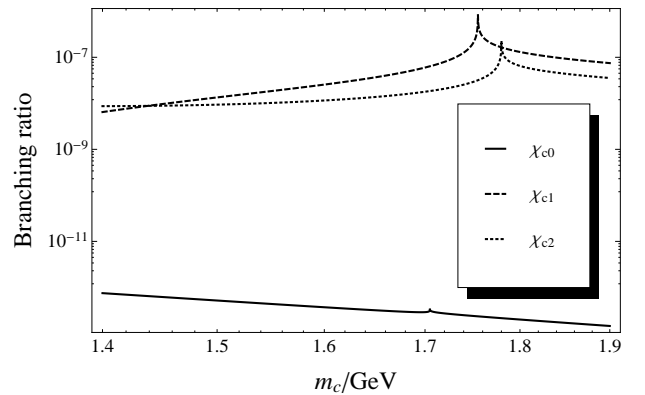}}
\caption{$\text{Br}[\chi_{cJ}\to e^+e^-]$ as functions of $m_c$.
}\label{ratiom}
\end{figure}

To show the dependence of our results on the charm quark mass, we
also plot $\text{Br}[\chi_{cJ}\to e^+e^-]$ as functions of $m_c$ in
 Fig.\ref{ratiom}.  The curves show the strong dependence of the branching ratios on $m_c$, or more precisely the binding energy $\omega$, which is mainly due to the $\ln m_c/\omega$ term in the decay amplitudes. And the cusps of the curves also originate from the discontinuity of this logarithmic term.

The most of $\chi_{cJ}$ events produced at BEPC-II are from the
radiative decays  $\psi(2S)\to \chi_{cJ}+\gamma$. BEPC-II will
produce about $3.2\times 10^9$ $\psi(2S)$ per year.  With
$\text{Br}[\psi(2S)\to \chi_{c0}+\gamma]=9.62\%$,
$\text{Br}[\psi(2S)\to \chi_{c1}+\gamma]=9.2\%$ and
$\text{Br}[\psi(2S)\to \chi_{c2}+\gamma]=8.74\%$, we expect that
there will be about $2.89\times10^8~\chi_{c0}$,
$2.76\times10^8~\chi_{c1}$ and $2.62\times10^8~\chi_{c2}$ produced
at BEPC-II annually. From the results listed in Table
\ref{Table:chic:r}, it seems that there is no possibility to see
$\chi_{c0}\to \ell^+\ell^-$, but marginal possibility to see
$\chi_{c1,2}\to \ell^+\ell^-$ at the BES-III experiments.

\subsubsection{$\chi_{bJ}\to\ell^+\ell^-$}
In our calculations, we take the following values for the masses of
$\chi_{bJ}$
\begin{eqnarray}
&&M_{\chi_{b0}}=9.86\,\text{GeV}\,,~~M_{\chi_{b1}}=9.89\,\text{GeV}\,,\nonumber\\
&&M_{\chi_{b0}}=9.91\,\text{GeV}\,.\end{eqnarray} The total decay
widths of the $\chi_{bJ}$ are absent in PDG review. $\chi_{bJ}$
decays mainly through annihilations to light hadrons (LH), or
radiative radiative transition to $ \Upsilon\gamma$ . Although the
branching ratios of $\chi_{bJ}\to \Upsilon\gamma$ have been measured
experimentally, and the corresponding decay widths can be calculated
within certain theoretical models, we will not use those
informations to infer the total decay width of $\chi_{bJ}$, since
both experimental measurements and theoretical calculations are not
so reliable at the present. Instead, we define the ratio between the
di-leptonic decay width and light hadronic decay width
\begin{align}
\mathcal{R}[\chi_{bJ}\to\ell^+\ell^-]=\frac{\Gamma[\chi_{bJ}\to\ell^+\ell^-]}{\Gamma[\chi_{bJ}\to\text{LH}]},
\end{align}
to get rough estimates on the orders of magnitudes of the branching
ratios of the di-leptonic decays, since the annihilations to light
hadrons dominate the decays of $\chi_{bJ}$.

With the decay widths $\Gamma[\chi_{bJ}\to \text{LH}]$ given in
\cite{Bodwin:1994jh},  we have
\begin{subequations}
\begin{align}
\mathcal{R}[&\chi_{b0}\to\ell^+\ell^-]=\nonumber\\
&\frac{3e^4_b\alpha^4 M^4_{\chi_{b0}}}{8\pi^2\alpha^2_S
m^4_b}\beta^3(1-\beta^2)\frac{|f_0|^2}{16+n_f
M^2_{\chi_{bJ}}\rho},\\
\mathcal{R}[&\chi_{b1}\to\ell^+\ell^-]=\nonumber\\
&\frac{\beta M^2_{\chi_{b1}}}{16\pi^2\alpha^2_S n_f
\rho}\left[\frac{4e^4_b \alpha^4}{m^4_b}|f_1|^2\beta^2+
G_F^2\left(2\beta^2
g^2_a+(3-\beta^2)g^2_v\right)\right.\nonumber\\
&\left. -\frac{4\sqrt{2} e^2_b
\alpha^2G_F}{m_b^2 }\beta^2g_a \text{Re}[f_1]\right],\\
\mathcal{R}[&\chi_{b2}\to\ell^+\ell^-]=\nonumber\\
&\frac{3 e^4_b\alpha^4 M^4_{\chi_{b2}}}{8\pi^2\alpha^2_S
m^4_b}\beta^3(5-2\beta^2)\frac{|f_2|^2}{64+15 n_f
M^2_{\chi_{bJ}}\rho},
\end{align}
\end{subequations}
where $n_f=4$ is the flavor number of light quarks,  $\alpha_S$ the
strong coupling constant, and
$\rho\equiv\frac{\langle\mathcal{O}_8(^1S_0)\rangle_{\chi_{bJ}}}{\langle\mathcal{O}_1(^3P_J)\rangle_{\chi_{bJ}}}$
which signifies the color-octet contributions to the annihilations
to light hadrons. We will use $\rho=0.0021~\text{GeV}^{-2}$ as in
\cite{Bodwin:matrix}. Thus, the hadronic uncertainties due to
$R^\prime(0)$ are greatly reduced in the ratio
$\mathcal{R}[\chi_{b0,2}\to\ell^+\ell^-]$ since
$M_{\chi_{b0,2}}^2\rho\sim 0.2$.  Meanwhile, $\Gamma[\chi_{b1}\to
\text{LH}]$ is dominated by the color-octet contribution, so
$\mathcal{R}[\chi_{b1}\to\ell^+\ell^-]$ is still sensitive to
parameter $\rho$.

In Table \ref{Table:chib:r}, we list $\mathcal{R}[\chi_{bJ}\to
\ell^+\ell^-]$ with $\alpha_s(M_{\chi_{bJ}}/2)=0.22$, $m_b=4.67$ GeV
and $\rho=0.0021~\text{GeV}^{-2}$.   One can see that the
$Z^0$-exchange dominates $\chi_{b1}\to \ell^+\ell^-$, and
$\mathcal{R}[\chi_{b1,2}\to \tau^+ \tau^-]$ deviates significantly
from $\mathcal{R}[\chi_{b1,2}\to e^+ e^-/\mu^+ \mu^-]$. This makes
our recalculations of the EM box diagrams meaningful.
\begin{table}[t]
\caption{\label{Table:chib:r}%
The ratio $\mathcal{R}[\chi_{bJ}\to \ell^+\ell^-]$ with $m_b=4.67$
GeV.}
\begin{center}
\begin{tabular}{lcc}
\hline\noalign{\smallskip}
Decay Channels  & QED  & QED+weak \\
\hline
$\chi_{b0}\to e^+e^-$    & $2.31\times10^{-14}$& \\
$\chi_{b0}\to\mu^+\mu^-$  & $6.80\times10^{-11}$& \\
$\chi_{b0}\to\tau^+\tau^-$  & $1.51\times10^{-9}$& \\
\hline
$\chi_{b1}\to e^+e^-$    & $3.29\times10^{-8}$    &$6.19\times10^{-7} $ \\
$\chi_{b1}\to\mu^+\mu^-$   & $3.27\times10^{-8}$  &$6.18\times10^{-7} $ \\
$\chi_{b1}\to\tau^+\tau^-$  & $2.04\times10^{-8}$ &$4.69\times10^{-7} $ \\
\hline
$\chi_{b2}\to e^+e^-$    & $4.18\times10^{-9}$& \\
$\chi_{b2}\to\mu^+\mu^-$   & $4.15\times10^{-9}$& \\
$\chi_{b2}\to\tau^+\tau^-$  & $2.54\times10^{-9}$& \\
\hline\noalign{\smallskip}
\end{tabular}
\end{center}
\end{table}

 To see the uncertainties of our results due to
the bottom quark mass $m_b$, we also list the values of
$\mathcal{R}[\chi_{bJ}\to \ell^+\ell^-]$ in Table \ref{Table:chib:m}
with several different values of $m_b$.


\begin{table}[h]
\caption{\label{Table:chib:m}$\mathcal{R}[ \chi_{bJ}\to
\ell^+\ell^-]$ with different values of $m_b$.}
\begin{center}
\begin{tabular}{lccc}
\hline\noalign{\smallskip}
{$m_b$}/GeV  &4.6      &4.8    &5.0    \\
\hline
$\chi_{b0}\to e^+e^-$     & $2.50\times10^{-14}$  & $2.02\times10^{-14}$ & $1.67\times10^{-14}$\\
$\chi_{b0}\to\mu^+\mu^-$  & $7.21\times10^{-11}$ & $6.24\times10^{-11}$  & $6.20\times10^{-11}$\\
$\chi_{b0}\to\tau^+\tau^-$  & $1.53\times10^{-9}$ & $1.55\times10^{-9}$  & $2.32\times10^{-9}$\\
\hline
$\chi_{b1}\to e^+e^-$    & $5.93\times10^{-7}$    &$6.94\times10^{-7} $ & $8.58\times10^{-7}$\\
$\chi_{b1}\to\mu^+\mu^-$   & $5.92\times10^{-7}$  &$6.93\times10^{-7} $ & $8.57\times10^{-7}$\\
$\chi_{b1}\to\tau^+\tau^-$  & $4.47\times10^{-7}$ & $5.29\times10^{-7}$  & $6.59\times10^{-7}$\\
\hline
$\chi_{b2}\to e^+e^-$    & $3.71\times10^{-9}$   & $5.76\times10^{-9}$   & $1.84\times10^{-8}$\\
$\chi_{b2}\to\mu^+\mu^-$   & $3.68\times10^{-9}$  & $5.73\times10^{-9}$  & $1.83\times10^{-8}$\\
$\chi_{b2}\to\tau^+\tau^-$  & $2.21\times10^{-9}$ & $3.63\times10^{-9}$  & $1.23\times10^{-8}$\\
\hline\noalign{\smallskip}
\end{tabular}
\end{center}
\end{table}


According to the estimations in \cite{Braguta}, the production cross
sections of $\chi_{bJ}(J=1,2)$ at LHC ($\sqrt{s}=14$TeV) are
\begin{align}
\sigma(p p\to\chi_{b0}+X)=1.5\mathrm{\mu b},~\sigma(p
p\to\chi_{b2}+X)=2\mathrm{\mu b},
\end{align}
so there are about $7.5\times 10^{9}\chi_{b0}$ and
$10^{10}\chi_{b2}$ produced in LHC, with present integrated
luminosity $5 \mathrm{fb}^{-1}$. Thus, there are about 11, 41 and 25
events of $\chi_{b0}\to \tau^+\tau^-$, $\chi_{b2}\to
e^+e^-(\mu^+\mu^-)$ and $\tau^+\tau^-$, respectively. Of course,
considering the efficiency of the detection, those di-leptonic
decays of $\chi_{bJ}$ are not measurable at LHC so far. However, the
long-term goal of LHC is to reach a integrated luminosity around
$3000\mathrm{fb}^{-1}$ by the end of LHC life. Then,  there will be
about $6.80\times 10^3$ events for $\chi_{b0}\to \tau^+ \tau^-$,
$2.51\times 10^4$ events for $\chi_{b2}\to e^+ e^-$,
$2.49\times10^4$ events for $\chi_{b2}\to \mu^+\mu^-$,
$1.52\times10^4$ events for $\chi_{b2}\to \tau^+\tau^-$ accumulated
at LHC, which make the dileptonic decay of $\chi_{b0,2}$ measurable.
So far, there is no estimation on the $\chi_{b1}$ production rate at
LHC. Assuming the production rate of $\chi_{bJ}$ is proportional to
its decay width to light hadrons, we expect that $\chi_{b1}\to
\ell^+\ell^-$ has a comparable chance to be measured at LHC as
$\chi_{b0,2}\to \ell^+\ell^-$.

\subsection{Impacts from new physics}

We now consider the numerical impacts on $\chi_{QJ}\to \ell^+\ell^-$
from the new physics effects. A complete analysis with considering
all possible parameters and constrains to parameter space seems
rather complicated and deviate from the main topic of the paper. We
will perform our analysis by choosing some specific values for the
new physics parameters.

\subsubsection{$\chi_{b0}\to \ell^+\ell^-$ in Type-II 2HDM}
In general Type-II 2HDM, we have to consider three individual
parameters: $m_h$, $\tan\beta$ and $\alpha$.  Recently, the Higgs
mass has been excluded in a broad range of mass parameters.
Recently, both the CMS and ATLAS collaborations observed an excess
of Higgs-like events with 2-3 sigma around 125 GeV in $pp$
collisions at the Large Hadron Collider (LHC)
 at $\sqrt{s} = 7$ TeV \cite{:2012si,Chatrchyan:2012tx}. In the following analysis, we roughly take $m_h$ around 125 GeV,  although the experimentally observed events may not be referred to the lightest neutral Higgs  in 2HDM.
\begin{figure}[!htb]
\centerline{\includegraphics[width=8.5cm]{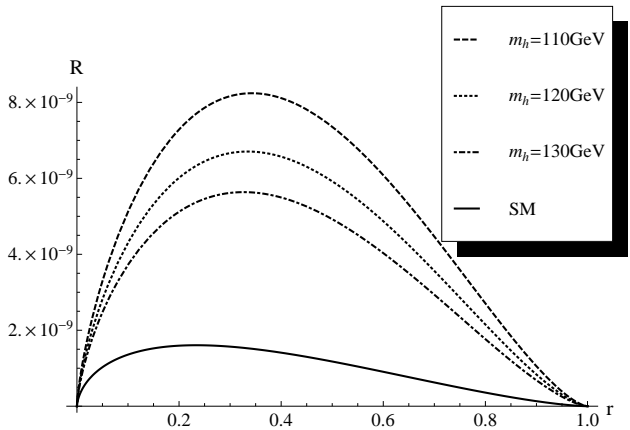}} \caption{
$\mathcal{R}[\chi_{b0}\to\ell^+\ell^-]$ as functions of
$r=m_\ell^2/m_b^2$ with  $\tan\beta=10$ and different choices of
$m_h$. }\label{mssmr}
\end{figure}

\begin{figure}[!htb]
\centerline{\includegraphics[width=9cm]{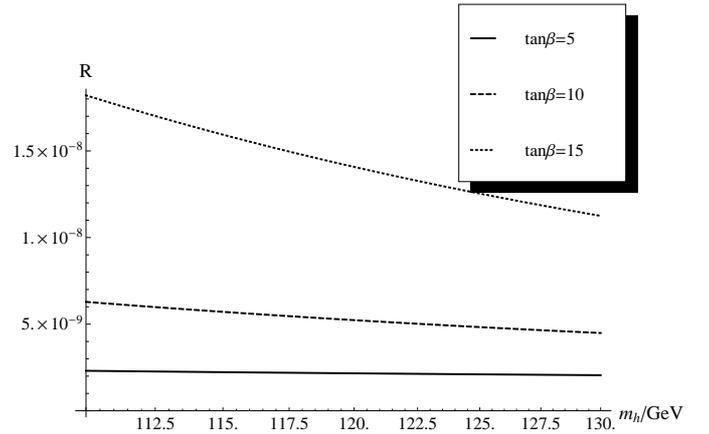}} \caption{
$\mathcal{R}[\chi_{b0}\to\ell^+\ell^-]$ as functions of $m_h$ with
different choices of $\tan\beta$. }\label{mssmm}
\end{figure}

We first take $\alpha\sim\beta$, while the mass of neutral Higgs
$m_h$ and $\tan\beta$ are free parameters. We plot
$\mathcal{R}[\chi_{b0}\to \ell^+\ell^-]$ as a function of
$r=m^2_\ell/m^2_b$ with $\tan\beta=10$ and different values of
$m_h$,  in Fig.\ref{mssmr}, and also plot $\mathcal{R}$ as a
function of $m_h$ with $\tan\beta=5\,,10\,,15$ in Fig. \ref{mssmm}.

It is easy to see that, when $\tan\beta=10$, the decay width is much
larger than the SM results. If $m_h$ is taken to be 125GeV,
$\mathcal{R}[\chi_{b0}\to \tau^+\tau^-]$ will be about
$4.83\times10^{-9}$, which is about three times of the SM prediction
$1.51\times10^{-9}$. When $\tan\beta=15$, the dileptonic decay will
be enhanced by 8 times. Thus, if Type-II 2HDM is a true theory of
our world and $\tan\beta$ is large, we may accumulate more
$\chi_{b0}\to\tau^+\tau^-$ events than the number predicted by SM at
LHC.

\begin{figure}[!htb]
\centerline{\includegraphics[width=9cm]{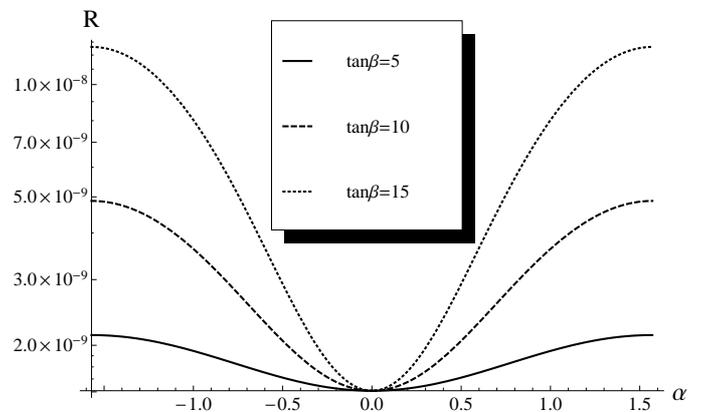}}
\caption{$\mathcal{R}[\chi_{b2}\to\ell^+\ell^-]$ as functions of
$\alpha$ with different choices of $\tan\beta$.
 }\label{mssmalpha}
\end{figure}

In general 2HDM, the neutral Higgs mixing angle $\alpha$ is not
strongly correlated with $\beta$. To see how
$\mathcal{R}[\chi_{b0}\to \ell^+\ell^-]$ depends on $\alpha$, we
plot $\mathcal{R}$ as a function of $\alpha$ with $\tan\beta=5,10$
and $15$, respectively at $m_h=125$GeV in Fig.\ref{mssmalpha}.
$\mathcal{R}[\chi_{b0}\to \tau^+\tau^-]$ reaches its minimum value
$1.51\times10^{-9}$ when $\alpha=0$ , which is just the SM result,
and it reaches its maximum value when $\alpha=\beta$, which is just
the case we have discussed above.

\subsubsection{$\chi_{b2}\to \ell^+\ell^-$ in the RS model}
In the RS model, the most important parameters are $\kappa\equiv
k/\overline{M}_{Pl}$. Recently, the ATLAS collaboration has reported
the 95\% C.L. lower limit on the mass of RS graviton for various
values of $k/\bar{M_{Pl}}$, which are 0.71,1.03,1.33,1.63TeV for
$k/\bar{M_{Pl}}=0.01,0.03,0.05,0.1$, respectively \cite{rslimit}.

\begin{figure}[h]
\centerline{\includegraphics[width=9cm]{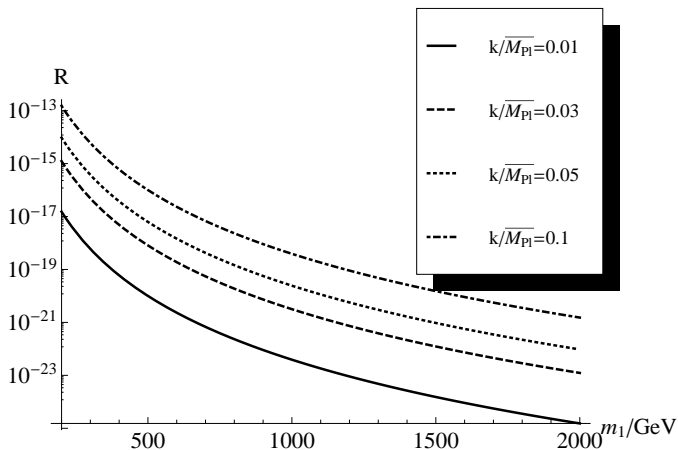}} \caption{
$\mathcal{R}[\chi_{b2}\to\ell^+\ell^-]$ as function of $m_1$ with
different choices of $k/\overline{M}_{\rm Pl}$. }\label{fig:rswidth}
\end{figure}

Here, we only consider the decay width induced by the lightest KK
excitation of the graviton with neglecting the interference between
graviton-exchange and SM contribution. It will not prevent us
getting a qualitative observation.  We plot the
$\mathcal{R}[\chi_{b2}\to \ell^+\ell^-]$ via the lightest KK
graviton as a function of $m_1$, with
$k/\overline{M}_{Pl}=0.01,0.03,0.05,0.1$ in Fig.\ref{fig:rswidth},
while the lepton mass $m_\ell$ is taken to be 0. From Fig.
\ref{fig:rswidth},  Table \ref{Table:chib:r} and \ref{Table:chib:m},
one can see that, even for $k/\overline{M}_{Pl}\sim 0.1$ and $m_1
\sim 200$ GeV, the KK graviton exchange cannot compete with the SM
contributions in $\chi_{b2}\to \ell^+\ell^-$. Thus, we have no
chance to look for any hints of the RS model in such decay.

\section{Summary\label{sect:sum}}

In this paper, we recalculate $\chi_{QJ}\to \ell^+\ell^-$ within the
SM by considering the finite mass of the leptons.  The suppression
of such decays in the SM make them sensitive to the NP. In the
future experiments where a huge number of quarkonia are produced,
such rare decays of quarkonia could be a new play-ground of NP
hunters other than high-energy collisions or flavor changing
processes. We investigate $\chi_{b0}\to \ell^+\ell^-$ in Type-II
2HDM, and $\chi_{b2}\to\ell^+\ell^-$ in the RS model. We find that
in the large $\tan\beta$ limit, we may have better chance to observe
$\chi_{b0}\to \tau^+\tau^-$ in Type-II 2HDM than in the SM, and no
chance to find the hints of the RS model in
$\chi_{b2}\to\ell^+\ell^-$. It could be interesting to investigate,
in which kind of extensions of the SM, the leptonic decays of
$\chi_{QJ}$ might be enhanced so that we can observe them.

\section*{Acknowlegement}
We thank Yu Jia and Chang-Zheng Yuan  for valuable discussions. This
work is partly supported by National Natural Science Foundation of
China under grant number 10705050 and 10935012.

\end{document}